\documentclass[10pt, prd, twocolumn, nofootinbib,preprint,superscriptaddress]{revtex4}
\pdfoutput=1
\usepackage[T1]{fontenc}
\usepackage{amsmath,amssymb}
\usepackage{epsfig}
\usepackage{graphicx}
\usepackage[usenames,dvipsnames]{color}
\usepackage{subfigure}
\usepackage{slashed}
\usepackage[colorlinks,citecolor=blue]{hyperref}
\usepackage{pdfpages}
\usepackage{color}
\usepackage{tikz} 
\usetikzlibrary{arrows,scopes} 

\usepackage{graphicx}
\usepackage{dcolumn}
\usepackage{bm}
\usepackage{hyperref}
\usepackage{bigints}
\usepackage{braket}

\begin{document}


\title{Constraining deviations from spherical symmetry using $ \gamma $-metric}

\author{Hrishikesh~Chakrabarty}
\email{hrishikeshchakrabarty@ucas.ac.cn}
\affiliation{School of Astronomy and Space Sciences, University of Chinese Academy of Sciences (UCAS), Beijing China}

\author{Yong Tang}
\email{tangy@ucas.ac.cn}
\affiliation{School of Astronomy and Space Sciences, University of Chinese Academy of Sciences (UCAS), Beijing China}
\affiliation{School of Fundamental Physics and Mathematical Sciences, Hangzhou Institute  for  Advanced  Study,  UCAS,  Hangzhou  310024,  China}
\affiliation{International  Center  for  Theoretical  Physics  Asia-Pacific,  Beijing/Hangzhou,  China}
\affiliation{National  Astronomical  Observatories,  Chinese  Academy  of  Sciences,  Beijing,  China}

\date{\today}

\begin{abstract}

The $ \gamma $-spacetime metric is a static and axially symmetric vacuum solution of the Einstein equation. This spacetime represents a naked singularity and it has an extra parameter $ \gamma $ which signifies deviations from spherical symmetry. In this work, we study the possibility of constraining the deformation parameter with astrophysical observations. We start with gravitational lensing in the weak and strong-field limits and calculate the respective deflection angles to show that only strong-field lensing observations will be able to constrain $ \gamma $ independently. Later we study two other classical tests of gravity: Shapiro time delay and precession of perihelion. We show that, out of these two experiments, the deformation parameter affects the observables only in perihelion shift.

\end{abstract}

\maketitle


\section{Introduction}\label{sec-intro}

The $ \gamma $-metric, also known as Zipoy-Voorhees spacetime\cite{gamma1,gamma2} is an exact vacuum solution of Einstein's field equations. The metric belongs to the Weyl class of spacetimes and is static, axially symmetric and asymptotically flat. In Erez-Rosen coordinates \cite{erez-rosen} the spacetime is given by the line element
\begin{equation}\label{e-gamma}
    \begin{aligned}
        {\rm d}s^2 = &- f^\gamma dt^2 + f^{\gamma^2-\gamma}g^{1-\gamma^2} \left( \frac{dr^2}{f} + r^2 d\theta^2 
        \right) \\
        &+ f^{1-\gamma} r^2\sin^2\theta d\phi^2,
    \end{aligned}
\end{equation}
where
\begin{equation}
    \begin{aligned}
        & f = 1 - \frac{2M}{r}, \\
        & g = 1 - \frac{2M}{r} + \frac{M^2\sin^2\theta}{r^2}.
    \end{aligned}
\end{equation}
There are two parameters that characterize the spacetime: $ M > 0 $ is related to the mass of source and $ \gamma > 0 $ quantifies deformation from spherical symmetry as for $\gamma > 1$ ($\gamma < 1$), the spacetime is oblate (prolate). The spacetime is spherically symmetric for $\gamma = 1$, and it corresponds to Schwarzschild solution. The total Arnowitt-Deser-Misner (ADM) mass measured by an observer at infinity is $ M_{\rm ADM} = \gamma M $.  

The metric contains a genuine curvature singularity at $ r = 2M $ for $ \gamma \neq 1 $, meaning the spacetime is a naked singularity~\cite{Virbhadra:1996cz,Papadopoulos:1981wr}. However, this surface at $ r = 2M $ must be regarded as an infinitely redshifted one which may show features similar to the Schwarzschild event horizon observationally \cite{Abdikamalov:2019ztb}. For this reason, the $ \gamma$-metric can be considered a ``black hole mimicker.'' It was shown in~\cite{Chakrabarty:2017ysw} that the singularity at $ r = 2M $ can be resolved in theories of conformal gravity leaving only the infinitely redshifted surface. 

Geodesic motion of massive and massless particles in this spacetime was studied in \cite{Herrera:1998eq,Chowdhury:2011aa,Boshkayev:2015jaa,Toshmatov:2019bda}.The $ \gamma $-metric has also been used to describe the exterior of a general relativistic disk, for details see \cite{Gonzalez:2003fe}. Recently, oscillation of neutrinos and its lensing was studied in $\gamma$-metric and it was shown that constraints on $ \gamma $ can be obtained from the detection of extra-solar neutrinos. There have been studies on optical properties and shadows of $ \gamma $-metric which verified the ``black hole mimicking'' property of the spacetime \cite{Abdikamalov:2019ztb,Shaikh:2021cvl}.  However, the $\gamma$-metric is non-integrable in general, as shown in \cite{Lukes-Gerakopoulos:2012qpc} and leads to interesting chaotic behavior for the motion of test particles. The fact that $ \gamma $-metric mimics a black hole and the particle motion scenario is very much similar to that in Schwarzschild spacetime enables it as a well-motivated and simple candidate to study toy models of astrophysical scenarios where the exterior of a compact massive compact object is not given by a usual black hole line element but by the deformed metric.

Despite this interesting motive, there have not been many efforts to study different types of experiments in $ \gamma $-spacetime. In this work, we discuss the possibility of constraining the $ \gamma $ parameter with astrophysical observations in the weak and strong field limits. We start with the calculation of the deflection angle in $\gamma $-metric for gravitational lensing in the strong and weak-field limit. The weak lensing deflection angle can be directly applied to solar-system experiments. However, we show that the parameter $ \gamma $ cannot be independently constrained with weak lensing observations. On the other hand, the strong lensing observations would be able to constrain $ \gamma $ with future very long baseline interferometry (VLBI) projects. Finally, we shift our attention to other two solar-system experiments: radar echo delay and precession of perihelia.

The rest of the paper is organized as follows: In Sec.~\ref{sec-gl}, we study the strong and weak field limit of gravitational lensing in $ \gamma $-metric. Then in sec.~\ref{sec-sd} and Sec.~\ref{sec-pp} we investigate the Shapiro time-delay and precession of perihelia in $ \gamma $-metric. Finally, we summarize and discuss our findings in Sec.~\ref{sec-od}. Throughout the paper we use natural units setting $ G = c = 1 $.

\section{Gravitational lensing}\label{sec-gl}

In this section, we would calculate the angle of deflection of light in $ \gamma $-metric and see if any observational constraints can be obtained. First, we shall work in the weak field to find an analytical expression for the deflection angle. Later, we extend our calculations to the strong field limit using a method first described in~\cite{Bozza:2002zj,Bozza:2010xqn}. 

\subsection{Weak-field limit}

The world lines of light rays in a curved spacetime can be described by giving the coordinates $ x^\alpha $ as functions of any one of a family of affine parameters $ \lambda $. Now, the null vector $ v^\alpha $ is a tangent to the world line given by
\begin{equation}
    v^\alpha \equiv \frac{dx^\alpha}{d\lambda}.
\end{equation}
The $ \gamma $-metric is independent of $ t $ and $ \phi $ coordinates and hence we shall have two conserved quantities along the light ray trajectories,
\begin{equation}\label{eq-el}
    \begin{aligned}
        e &\equiv -\boldsymbol\varepsilon . \mathbf{v} = f^\gamma\left(\frac{dt}{d\lambda}\right) \Rightarrow \left(\frac{dt}{d\lambda}\right) = \frac{e}{f^\gamma}, \\
        l &\equiv \boldsymbol\eta . \mathbf{v} = r^2 f^{1-\gamma} \left( \frac{d\phi}{d\lambda} \right) \Rightarrow \left( \frac{d\phi}{d\lambda} \right) = \frac{l}{r^2f^{1-\gamma}}, 
    \end{aligned}
\end{equation}
where $ \eta^\alpha = (0,0,0,1) $ and $ \varepsilon^\alpha = (1,0,0,0) $ are Killing vectors of the field. A third integral can be written considering the requirement that the tangent vector be null
\begin{equation}
    \mathbf{v} . \mathbf{v} = g_{\alpha\beta}\frac{dx^\alpha}{d\lambda}\frac{dx^\beta}{d\lambda} = 0.
\end{equation}
Now, on the equatorial plane ($ \theta = \pi/2 $), the previous equation can be written explicitly as
\begin{equation}
    f^\gamma \left( \frac{dt}{d\lambda} \right)^2 - \frac{g^{1-\gamma^2}}{f^{1+\gamma-\gamma^2}}\left( \frac{dr}{d\lambda} \right)^2 - f^{1-\gamma}r^2\left( \frac{d\phi}{d\lambda} \right)^2 = 0.
\end{equation}
From Eq.~\eqref{eq-el}, we replace $ (dt/d\lambda) $ and $ (d\phi/d\lambda) $ to obtain
\begin{equation}
    \frac{e^2}{f^\gamma} - \frac{g^{1-\gamma^2}}{f^{1+\gamma-\gamma^2}}\left( \frac{dr}{d\lambda} \right)^2 - f^{\gamma-1}\left(\frac{l^2}{r^2}\right) = 0.
\end{equation}
Multiplying by $ (f^\gamma/l^2) $, it gives
\begin{equation}\label{eq-drdlambda}
    \frac{1}{u^2} = \frac{1}{l^2}\left[ f^{\gamma^2-1}g^{1-\gamma^2} \right]\left( \frac{dr}{d\lambda} \right)^2 + W_{\rm eff}(r),
\end{equation}
where $ u = l/e $ is the impact parameter and 
\begin{equation}
    W_{\rm eff}(r) = \frac{1}{r^2}f^{2\gamma-1}
\end{equation}
is the effective potential felt by the photons. From Eq.~\eqref{eq-el}, we know
\begin{equation}\label{eq-dphidlambda}
    \left( \frac{d\phi}{d\lambda} \right)^2 = \left(\frac{l}{r^2f^{1-\gamma}}\right)^2.
\end{equation}
Dividing Eq.~\eqref{eq-dphidlambda} by Eq.~\eqref{eq-drdlambda}, we get
\begin{equation}\label{eq-dphidr}
    \left( \frac{d\phi}{dr} \right)^2 = \frac{f^{\gamma^2-1}g^{1-\gamma^2}}{r^4f^{2(1-\gamma)}\left( \frac{1}{u^2} - \frac{f^{2\gamma-1}}{r^2} \right)} = \mathcal{Z}.
\end{equation}
This equation gives us the change in azimuthal angle with respect to the radial coordinate and it can be integrated to obtain the deflection angle $ \Delta \phi $. The magnitude of the total angel swept out as the light ray proceeds in from infinity and back out again. $ \Delta \phi $ is just twice the angle swept out from a turning point $ r = r_1 $ to infinity. Therefore, 
\begin{equation}\label{eq-master-defang-eqplane}
    \Delta \phi = 2\int d\phi = 2\int^\infty_{r_{1}} \sqrt{\mathcal{Z}} dr,
\end{equation}
and the turning point $ r_1 $ is the radius where $ 1/u^2 = W_{\rm eff}(r_1) $. The integral here is very complicated and cannot be solved analytically. Hence, we would consider a weak-field limit where $ M/r << 1 $. To obtain an analytical solution, we introduce a new variable $ w $, such that
\begin{equation}\label{eq-cov}
    r = \frac{u}{w} \Rightarrow dr = -\frac{u}{w^2}dw.
\end{equation}
With the new variable, the functions of the metric coefficients in the weak-field limit,
\begin{equation}\label{eq-wfl}
    f \simeq g \simeq \left( 1 - \frac{2M}{u}w \right)
\end{equation}
Now using Eq.~\eqref{eq-cov} and Eq.~\eqref{eq-wfl} in the integral, we obtain
\begin{equation}\label{wl-int}
    \delta \phi = 2\int^{w_1}_{0} \frac{1+\frac{M}{u}w}{\left[ 1 + (2\gamma-1)\frac{2M}{u}w - w^2 \right]^{1/2}} dw,
\end{equation}
where the limit $ w_1 $ is the value of $ w $ at which the denominator vanishes. This integral can be solved analytically using a standard algebraic manipulation software. The result is
\begin{equation}
    \begin{aligned}
        \Delta \phi = &2 \Bigg[ -a_1\sqrt{1+a_1w-w^2} \\
        &- \arctan \left( \frac{a_2-2w}{2\sqrt{1+a_1w-w^2}} \right)\Bigg]_0^{w_1},
    \end{aligned}
\end{equation}
where $ a_1 = M/u $ and $ a_2 = (2\gamma-1)(2M/u) $. Taking the limit, we get
\begin{equation}
    \Delta \phi = \pi + \frac{4\gamma M }{u}
\end{equation}
The deflection angle $ \delta \phi_{\rm def} $ is related to $ \Delta \phi $ by the following relation
\begin{equation}
    \delta \phi_{\rm def} = \Delta\phi -\pi.
\end{equation}
Thus, the deflection angle in the weak field limit is 
\begin{equation}
    \delta \phi_{\rm def}^\gamma = \frac{4\gamma M}{u} = \frac{4M_{\rm ADM}}{u}.
\end{equation}
Now, we can see from the deflection angle that even if the expression involves the parameter $ \gamma $, it is unlikely  to constrain it by observing gravitational lensing in the weak-field limit. The gravitational mass of the system is $ M_{\rm ADM} = \gamma M $ and the observations will only provide information about it, unable to break the degeneracy. In the following section, we shall see if it is possible to break this degeneracy in the strong field limit.

\subsection{Strong field limit: Bozza's method}

\begin{figure*}[]
    \begin{center}
        \includegraphics[width=8cm]{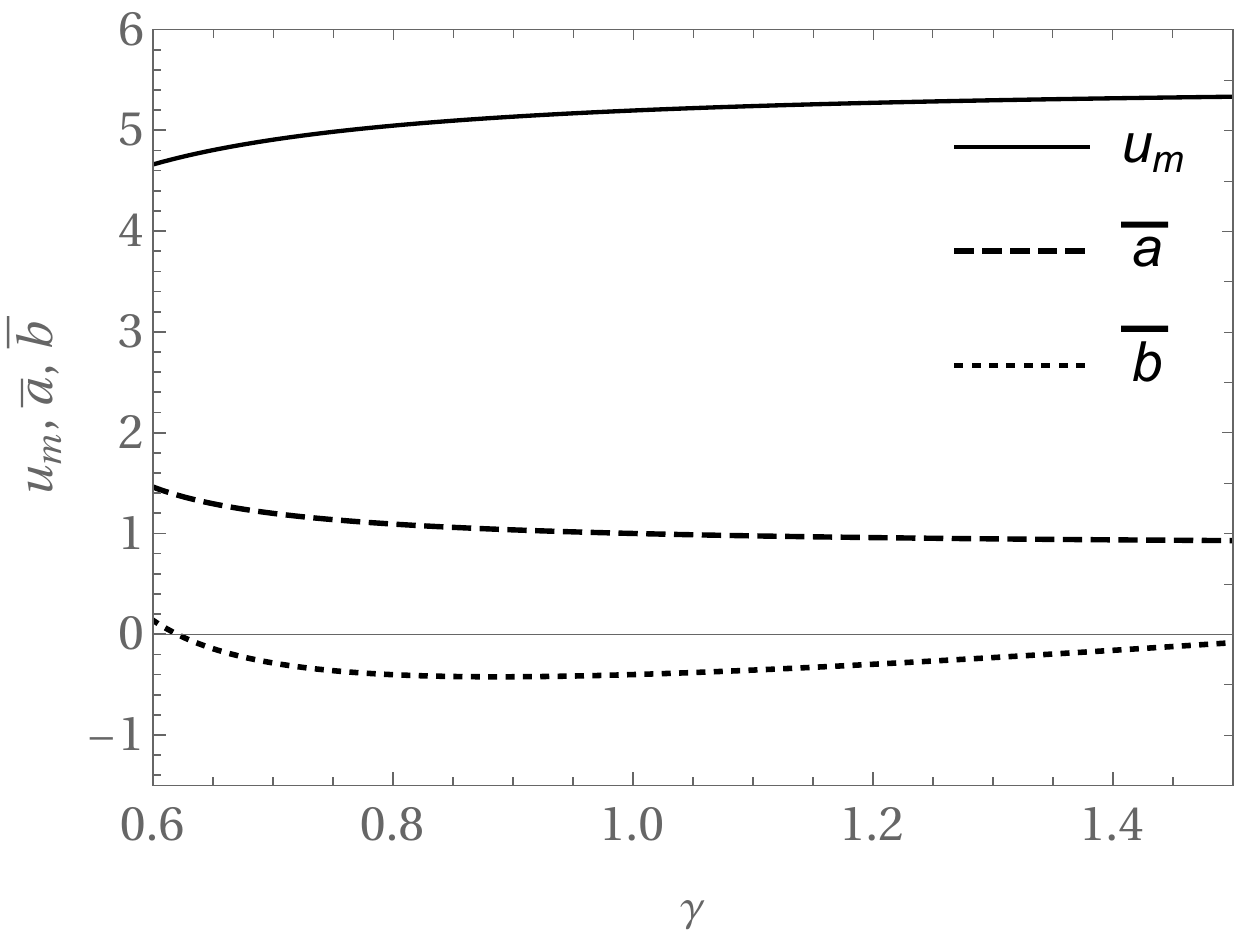}
        \includegraphics[width=8cm]{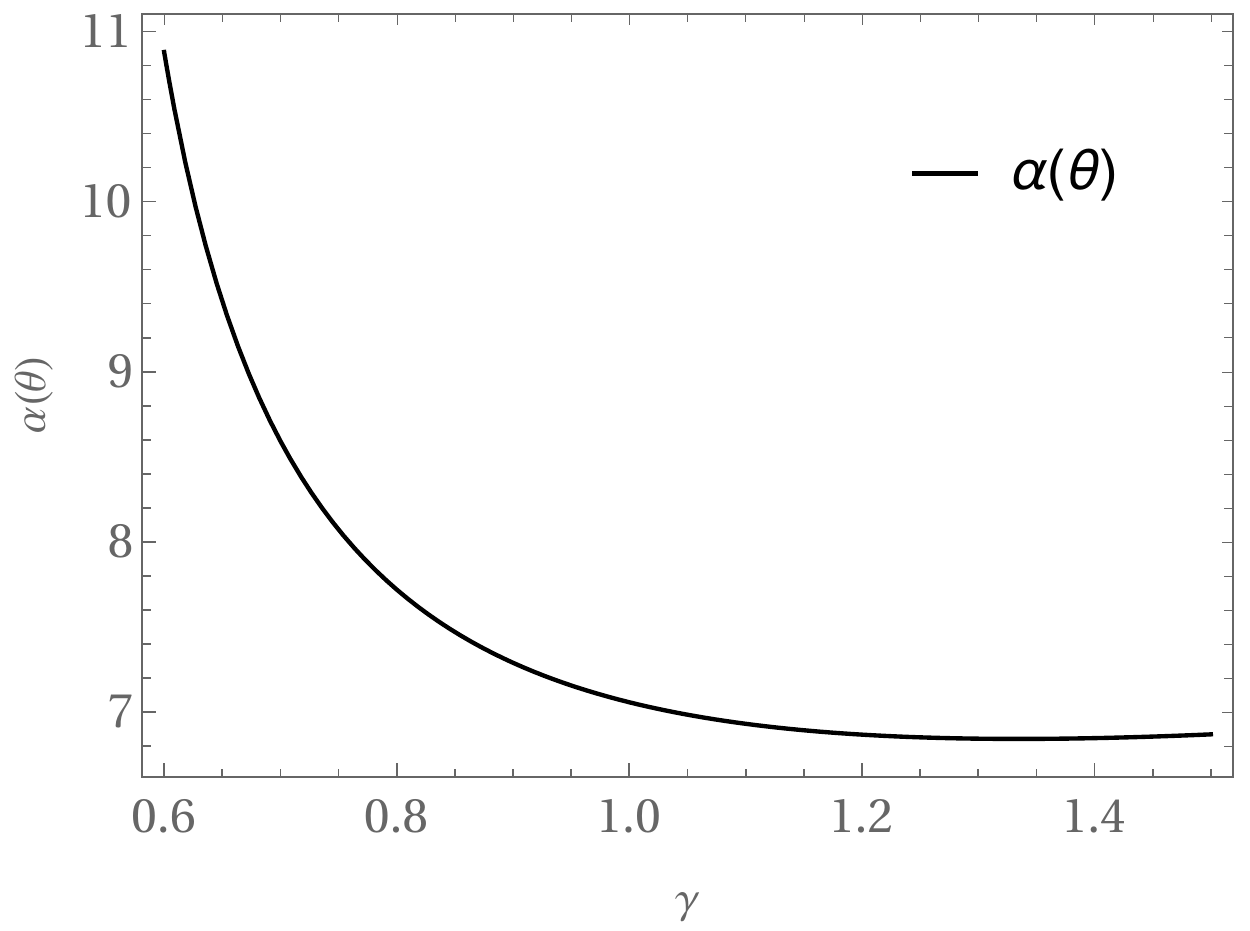}
    \end{center}
    \vspace{-0.5cm}
    \caption{Left panel: Coefficients of the deflection angle (Eq.~\eqref{sl-def-ang}) $ \Bar{a}, \Bar{b} $ and $ u_m $ with respect to $ \gamma $ evaluated at $ u - u_m = 0.003 $. Right panel: The deflection angle $ \alpha(\theta) $ with respect to $ \gamma $ evaluated at $ u - u_m = 0.003 $. \label{fig-sl}}
\end{figure*}

\begin{table*}[t]
\centering
  \begin{tabular}{ |p{3cm}||p{2.2cm}|p{2.2cm}|p{2.2cm}|p{2.2cm}|p{2.2cm}|  }
 \hline
  Parameters & $ \gamma = $ 0.95 & $ \gamma = $ 0.97 & $ \gamma = $ 1.0 (Schwarzschild) & $ \gamma = $ 1.02 & $ \gamma = $ 1.05 \\
 \hline
 $ \theta_\infty (\mu{\rm arcsec}) $ & 16.81 & 16.85 & 16.9 & 16.94 & 16.98 \\
 $ s (\mu{\rm arcsec}) $ & 0.02303 & 0.0221 & 0.0211 & 0.02057 & 0.01984  \\
 $ p_m $ (magnitude) & 6.714 & 6.758 & 6.821 & 6.859 & 6.912 \\
 $ u_m/R_s $ & 2.58 & 2.589 & 2.598 & 2.603 & 2.609 \\
 $ \Bar{a} $ & 1.0159 & 1.0091 & 1.0 & 0.9944 & 0.9869 \\
 $ \Bar{b} $ & -0.4148 & -0.4097 & -0.4002 & -0.3927 & -0.3801 \\
 \hline
\end{tabular}
\caption{Estimates of the observables and the coefficients $ \Bar{a}, \Bar{b} $ and $ u_m $ of strong lensing for the black hole at the center of our galaxy considering $ \gamma $-metric as the spacetime geometry of the exterior. $ \theta_\infty, s $ and $ p_m(p) = 2.5 \log p $ are defined in the text. Here $ R_s = \frac{2GM}{c^2} $ is the Schwarzschild radius. }
\label{table}
\end{table*}

In this subsection, we attempt to obtain an analytical expression for gravitational lensing in the strong field limit in $ \gamma $-metric in the equatorial plane. This analytical procedure was developed in~\cite{Bozza:2001xd} for Schwarzschild spacetime and later extended to general cases in \cite{Bozza:2010xqn}. Other works on gravitational lensing of black holes and naked singularities can be found in~\cite{Claudel:2000yi}-\cite{Virbhadra:2008ws}. We shall first describe the method briefly and then apply it to $ \gamma $-metric. For convenience, we shall keep the notation used in \cite{Bozza:2010xqn}. 

Let us first express the metric in the equatorial plane as 
\begin{equation}
    ds^2 = -A(r)dt^2 + B(r)dr^2 +C(r)d\phi^2,
\end{equation}
where $ A(r) $, $ B(r) $ and $ C(r) $ are metric coefficients. To check the existence of photon sphere around a compact object described by $ \gamma $-metric, we require that the equation \cite{Virbhadra:2002ju,Claudel:2000yi}
\begin{equation}\label{psphere}
    \frac{C'(r)}{C(r)} = \frac{A'(r)}{A(r)}
\end{equation}
admits at least one positive solution and the largest root of this equation will be called photon sphere, $ r_m $. Also note that $ A, B, C, A' $ and $ C' $ must be positive for $ r > r_m $.  

The strong field expansion takes the photon sphere as the starting point. For $ \gamma $-metric, the photon sphere is at $ r_m = (2\gamma + 1)M $ \cite{Abdikamalov:2019ztb}. Note that, when $ \gamma \neq 1 $, the coordinates are not spherical and hence photon sphere will not be a sphere. However, in the equatorial plane, this issue does not make any difference and the strong lensing method can be applied to $ \gamma $-spacetime.   

A photon coming from infinity with an impact parameter $ u $ will be deviated while approaching the compact object, reaching a minimum distance $ r_0 $ and then emerging in another direction. The approach phase will be symmetrical to the departure phase with the time reversed. Now, using the conservation of angular momentum, the closest approach distance can be related to the impact parameter $ u $ by the equation 
\begin{equation}
    u = \sqrt{\frac{C_0}{A_0}},
\end{equation}
where the subscript $ 0 $ indicated that the functions are evaluated at $ r_0 $. Similar to the previous section, the geodesic equation can be easily used to extract the quantity 
\begin{equation}
    \frac{d\phi}{dr} = \frac{\sqrt{B}}{\sqrt{C}\sqrt{\frac{C}{C_0}}\frac{A_0}{A}-1},
\end{equation}
which gives the angular shift of the photon as a function of the radial coordinate. 
Now, in the strong field limit, the deflection angle can be calculated as a function of the closest approach \cite{Virbhadra:1998dy}
\begin{equation}
    \alpha(r_0) = I(r_0) - \pi,
\end{equation}
where
\begin{equation}\label{sl-int}
    I(r_0) = \int_{r_0}^\infty \frac{\sqrt{B}}{\sqrt{C}\sqrt{\frac{C}{C_0}\frac{A_0}{A}-1}}dr.
\end{equation}
We can see that for a vanishing gravitational field, $ \alpha (r_0) $ vanishes identically. In the weak-field limit, the integrand can be expanded to the first order in the gravitational potential reproducing Eq.~\eqref{wl-int}. Decreasing the impact parameter increases the deflection angle and at some point, it will exceed $ 2\pi $ resulting in a loop around the compact object spacetime. $ r_0 = r_m $ corresponds to the impact parameter $ u = u_m $, and at this point the deflection angle diverges meaning the photon gets captured.   

The divergence here is logarithmic~\cite{Bozza:2001xd,Bozza:2010xqn}. The analytical expansion for the deflection angle close to the divergence has the form 
\begin{equation}
    \alpha (r_0) = -a \log \left( \frac{r_0}{r_m} - 1 \right) + b + \mathcal{O}(r_0 - r_m),
\end{equation}
where the coefficients depend on the metric functions evaluated at $ r_m $. 

The deflection angle, $ \alpha(r_0) $ can be expressed as a function of the angular separation of the image from the lens, $ \theta $. Angular separation of the lens from the image is defined as $ \theta = {u}/{D_{OL}} $, where $ D_{OL} $ is the distance between the lens and the observer. In terms of this variable the deflection angle is expressed as 
\begin{equation}
    \alpha(\theta) = -\Bar{a}\log\left( \frac{\theta D_{OL}}{u_m} - 1 \right) + \Bar{b} + \mathcal{O}(u - u_m). 
\end{equation}
The coefficients $ \Bar{a} $ and $ \Bar{b} $ are as follows
\begin{equation}\label{sl-def-ang}
    \begin{aligned}
        \Bar{a} &= \frac{a}{2} = \frac{R(0,r_m)}{2\sqrt{\beta_m}}, \\
        \Bar{b} &=  -\pi + b_R + \Bar{a}\log \left(\frac{2\beta_m}{y_m}\right).
    \end{aligned}
\end{equation}
Here, $ R(0,r_m) $, $ \beta_m $, $ y_m $ are again functions of metric coefficients evaluated at $ r_0 = r_m $. The derivation of the deflection angle along with the forms of the coefficients in the strong field limit is rather long and is omitted in the main text. For reference we have added in Appendix \ref{appA} the derivation, which alternatively can also be found in \cite{Bozza:2010xqn}. Below, we apply this procedure to $ \gamma $-metric.

Our task is to find the coefficients $ \Bar{a} $, $ \Bar{b} $ and $ u_m $ to check how the deflection angle varies with respect to the parameter $ \gamma $. We rewrite the metric coefficients again as $ A, B $ and $ C $ for $ \gamma $-metric
\begin{equation}
    \begin{aligned}
    &A(r) = \left( 1 - \frac{2M}{r} \right)^\gamma, \\
    &B(r) = \left( 1 - \frac{2M}{r} \right)^{\gamma^2 -\gamma -1}\left( 1 - \frac{2M}{r} + \frac{M^2}{r^2}\right)^{1 - \gamma^2}, \\
    &C(r) = \left( 1 - \frac{2M}{r} \right)^{1-\gamma}r^2.
    \end{aligned}
\end{equation}
We can calculate the radius of the photon sphere by solving the equation $ \alpha = 0 $ (first equation of \eqref{alphabeta} of Appendix~\ref{appA}). This yields
\begin{equation}
    r_m = (2\gamma + 1)M. 
\end{equation}
Now, $ \beta $ (Eq.~\eqref{betam}) at $ r_0 = r_m $ is 
\begin{equation}
    \beta_m = \frac{(2 \gamma -1) (2 \gamma +1) \left[\left(\frac{2\gamma-1}{2 \gamma +1}\right)^{\gamma }-1\right]^2}{4 \gamma ^2\left(\frac{2\gamma-1}{2 \gamma +1}\right)^{\gamma }}.
\end{equation}
The integral Eq.~\eqref{bR} cannot be solved exactly to find $ b_R $. Therefore we expand the integrand in powers of $ (\gamma - 1) $ and evaluate the single coefficients. We get 
\begin{equation}
    b_R = b_{R,0} + b_{R,1}(\gamma - 1) + \mathcal{O}(\gamma - 1)^2.
\end{equation}
Here $ b_{R,0} $ is the value of the coefficient for Schwarzschild spacetime \cite{Bozza:2010xqn}
\begin{equation}
    b_{R,0} = 0.9496.
\end{equation}
$ b_{R,1} $ represents the correction term for $ \gamma $-metric. [see Appendix for expression].

Finally, we can compute the coefficients of lensing in the strong field limit. For $ M_{\rm ADM} = \gamma M  = 1 $,

\begin{widetext}
    \begin{equation}
        \begin{aligned}
        \Bar{a} &= \sqrt{\left(\frac{4\gamma^2}{4 \gamma^2 -1}\right)^{1- \gamma ^2}}, \\
        \Bar{b} &= -\pi + b_R + \sqrt{\left(\frac{4\gamma^2}{4 \gamma^2 -1}\right)^{1- \gamma ^2}}\log \left(\left(2 -\frac{1}{2 \gamma ^2}\right)  \left[1-\left(\frac{2 \gamma -1}{2 \gamma +1}\right)^{-\gamma}\right]^2\right), \\
        u_m &= \left(2 +\frac{1}{\gamma }\right) \left(\frac{2 \gamma -1}{2 \gamma +1}\right)^{\frac{1-2\gamma}{2}}.
        \end{aligned}
    \end{equation}
\end{widetext}

These parameters are functions of $ \gamma $ only, since we have fixed the observable mass. Now we can calculate the dependence of the deflection angle on the parameter $ \gamma $. In Schwarzschild spacetime, it was shown in \cite{Bozza:2010xqn} that the most external image appears where $ \alpha(\theta) $ falls below $ 2\pi $ and this happens at the location $ u - u_m = 0.003264 $. To see the dependence of the strong lensing parameters and deflection angle on $ \gamma $, we evaluate them at $ u = u_m + 0.003 $ and plot the result in Fig.~\ref{fig-sl}. 

We can see from the left panel of Fig.~\ref{fig-sl} that $ u_m $ increases while $ \Bar{a} $ decreases with increasing $ \gamma $. $ \Bar{b} $ on the other hand initially decreases with increasing $ \gamma $ reaching a minimum value at $ \gamma = 0.89 $ and later again increases. The panel shows the decrease of deflection angle $ \alpha(\theta) $ with increasing $ \gamma $. Both these plots show an asymptotic behaviour when we approach $ \gamma \rightarrow 0.5 $. This is precisely because of the disappearance of the photon sphere at $ \gamma = 0.5 $ as the photon sphere radius becomes $ r_m = 2M $ and it coincides with the singularity which is also an infinitely redshifted surface at the equatorial plane. 

Let us now have a look at one realistic numerical example. We have significant evidence that the center of our galaxy hosts a supermassive black hole with mass $ M_{\rm BH} \simeq 4 \times 10^6 {\rm M_\odot} $. Gravitational lensing by such a black hole was discussed in detail by Virbhadra and Ellis in \cite{Virbhadra:1999nm} for the Schwarzschild case and Bozza in \cite{Bozza:2002zj} comparing modified spacetimes. Considering a distance of $ D_{OL} = 8.5 {\rm kpc} $ between the black hole and the sun, they showed that the separation between each set of relativistic images with respect to the central lens would be $ \theta_\infty = 17 \mu{\rm arcsecs} $. Now we are on the verge of achieving such a resolution through actual VLBI projects such as Event Horizon Telescope \cite{EventHorizonTelescope:2019uob}. Here we follow the same procedure~ \cite{Bozza:2002zj,Virbhadra:1999nm} for $ \gamma $-metric and estimate the quantities required for a complete strong field limit reconstruction.

The observables for gravitational lensing are discussed in \cite{Virbhadra:1999nm,Bozza:2002zj,Virbhadra:2002ju}. Basically we are interested in two quantities
\begin{equation}
    \begin{aligned}
    &s = \theta_1 - \theta_{\infty}, \\
    &p = \frac{\mu_1}{\sum_{n=2}^\infty \mu_n}.
    \end{aligned}
\end{equation}
Here, $ s $ and $ p $ represents the separation between the first image and the others and the ratio between the flux of the first image and that coming from the others respectively. Clearly, $ \theta_1 $ is the angular separation for the first image and $ \theta_\infty $ signify the asymptotic position by the other set of images. $ \mu_n $ is the flux, where the subscript represents the corresponding image. The parameters $ s $ and $ p $ can be related to the coefficients of strong lensing as
\begin{equation}
    \begin{aligned}
    &s = \theta_\infty \exp^{\frac{\Bar{b}}{\Bar{a}}-\frac{2\pi}{\Bar{a}}} \\
    &p = \exp^{\frac{2\pi}{\Bar{a}}}.
    \end{aligned}
\end{equation}
So, just by measuring the angular separation and a ratio of flux, one can reconstruct the full strong field limit expansion of the deflection angle. 

In Table~\ref{table}, we show the estimates of these parameters for different values of $ \gamma $, including $ \gamma = 1 $ which is the Schwarzschild case. Looking at the table, it is immediately clear that the easiest quantity to evaluate is the minimum impact parameter $ u_m $. Once we achieve a microarcsecond resolution in the next years, it would become possible to distinguish between a Schwarzschild black hole and a naked singularity represented by $ \gamma $-spacetime metric. On the other hand, to fit all other coefficients into any compact object model, we need to separate at least the outermost relativistic image from the others. This can be done only with an increased optical resolution at least by two orders of magnitude concerning actual observational projects \cite{Bozza:2002zj}.

\section{Shapiro time delay}\label{sec-sd}

Another interesting relativistic effect in the propagation of light rays is the apparent delay in propagation time for a light signal passing near the sun. This is important because radar-ranging techniques can measure this delay and give constraint on $ \gamma $-metric. This effect is called the Shapiro time delay effect.

The idea is to measure the time required for radar signals to travel to an inner planet or satellite in two circumstances: a) when the signal passes very near the Sun and b) when the ray does not go near the Sun. The time required to travel for light $ t_0 $ between two planets sitting far away from the Sun is given by 
\begin{equation}
    t_0 = \int_{-l_1}^{l_2} dy,
\end{equation}
where $ dy $ is the differential distance in the radial direction in the solar system and $ l_1 $ and $ l_2 $ are the distances of the the planets from the Sun. When the radar signal travels close to the sun, the previous equation should be modified as 
\begin{equation}
    t = \int_{-l_1}^{l_2} \frac{dy}{v} = \int_{-l_1}^{l_2} \sqrt{\frac{g_{rr}}{g_{tt}}}dy. 
\end{equation}
Here, $ v = \sqrt{g_{rr}/g_{tt}} $ is the speed of light in presence of the gravitational field, $ g_{tt} $ and $ g_{rr} $ are the metric components. Now the time difference is 
\begin{equation}\label{eq-td-int}
    \Delta t = t - t_0 = \int_{-l_1}^{l_2} \left( \sqrt{\frac{g_{rr}}{g_{tt}}} - 1 \right) dy.
\end{equation}
The radial coordinate can be expressed as $ r = \sqrt{R_{\odot}^2+y^2} $, where $ R_{\odot} $ is the radius of the Sun. Now Eq.~\eqref{eq-td-int} becomes an integral of variable $ y $, so we can write
\begin{equation}
    \Delta t = \int_{-l_1}^{l_2} \left( \sqrt{\frac{g_{rr}(\sqrt{R_{\odot}^2+y^2})}{g_{tt}(\sqrt{R_{\odot}^2+y^2})}} - 1 \right)dy.
\end{equation}
Now $ g_{rr} $ and $ g_{tt} $ can be replaced in the above integral
\begin{equation}
    \Delta t = \int_{-l_1}^{l_2} \left( \sqrt{f^{\gamma^2-2\gamma-1}g^{1-\gamma^2}} - 1 \right)dy
\end{equation}
Explicitly, in the equatorial plane
\begin{widetext}
\begin{equation}
    \Delta t = \bigints_{-l_1}^{l_2} \left( \sqrt{ \left( 1 - \frac{2M}{\sqrt{R_{\odot}^2+y^2}} \right)^{\gamma^2-2\gamma-1}  \left( 1 - \frac{2M}{\sqrt{R_{\odot}^2+y^2}} + \frac{M^2}{R_{\odot}^2+y^2} \right)^{1-\gamma^2} } - 1 \right)dy.
\end{equation}
\end{widetext}

Now the integration can be performed considering a Earth-Sun-Mars system. Relevant quantities here are the distance to the Earth from the Sun $ R_{\rm E} = l_1 = 1.525 \times 10^{13} {\rm cm} $, the distance to the Mars from the Sun $ R_{\rm M} = l_2 = 2.491 \times 10^{13} {\rm cm} $, mass of the Sun $ M_\odot =\gamma M= 1.989 \times 10^{33} {\rm g} $ and the radius of the Sun $ R_\odot = 6.955 \times 10^{10} {\rm cm} $. 

For Schwarzschild metric, with the above mentioned values of the parameters the radar echo delay has the value 
\begin{equation}
    \Delta t_{RD}^{(Sch)} \simeq 4M_\odot \ln (4l_1l_2/R_\odot^2) \simeq 2.4927 \times 10^{-4} {\rm s} 
\end{equation}

In $ \gamma $-spacetime, the radar echo delay expression can be written as
\begin{widetext}
\begin{equation}\label{eq-td-integral}
    \Delta t_{RD}^\gamma = 2\bigints_{-l_1}^{l_2} \left( \sqrt{ \left( 1 - \frac{2(M_\odot/\gamma)}{\sqrt{R_{\odot}^2+y^2}} \right)^{\gamma^2-2\gamma-1}  \left( 1 - \frac{2(M_\odot/\gamma)}{\sqrt{R_{\odot}^2+y^2}} + \frac{(M_\odot/\gamma)^2}{R_{\odot}^2+y^2} \right)^{1-\gamma^2} } - 1 \right)dy
\end{equation}
\end{widetext}

First let us check the weak field limit of the above integral. We expand the expression under the square root in terms of $ M/R_\odot $ and keep only the terms of order $ \mathcal{O}(M/R_\odot) $. The integral becomes
\begin{equation}
    \begin{aligned}
        \Delta t_{RD}^\gamma &= 2\int_{-l_1}^{l_2}\frac{2\gamma (M_\odot/\gamma)}{R_\odot}\left( 1-\frac{y^2}{R_\odot^2} \right)^{-1}dy \\
        &= 2\int_{-l_1}^{l_2}\frac{2M_\odot}{R_\odot}\left( 1-\frac{y^2}{R_\odot^2} \right)^{-1}dy \\
        &\simeq 4M_\odot \ln\left( \frac{4l_1l_2}{R_\odot^2} \right) = \Delta t_{RD}^{Sch},  
    \end{aligned}
\end{equation}
which is nothing but the expression for Shapiro time-delay in Schwarzschild spacetime. Therefore, the light rays in the time delay experiment will not experience the effect of $ \gamma $ on them. 

\begin{figure*}[]
    \begin{center}
        \includegraphics[width=12cm]{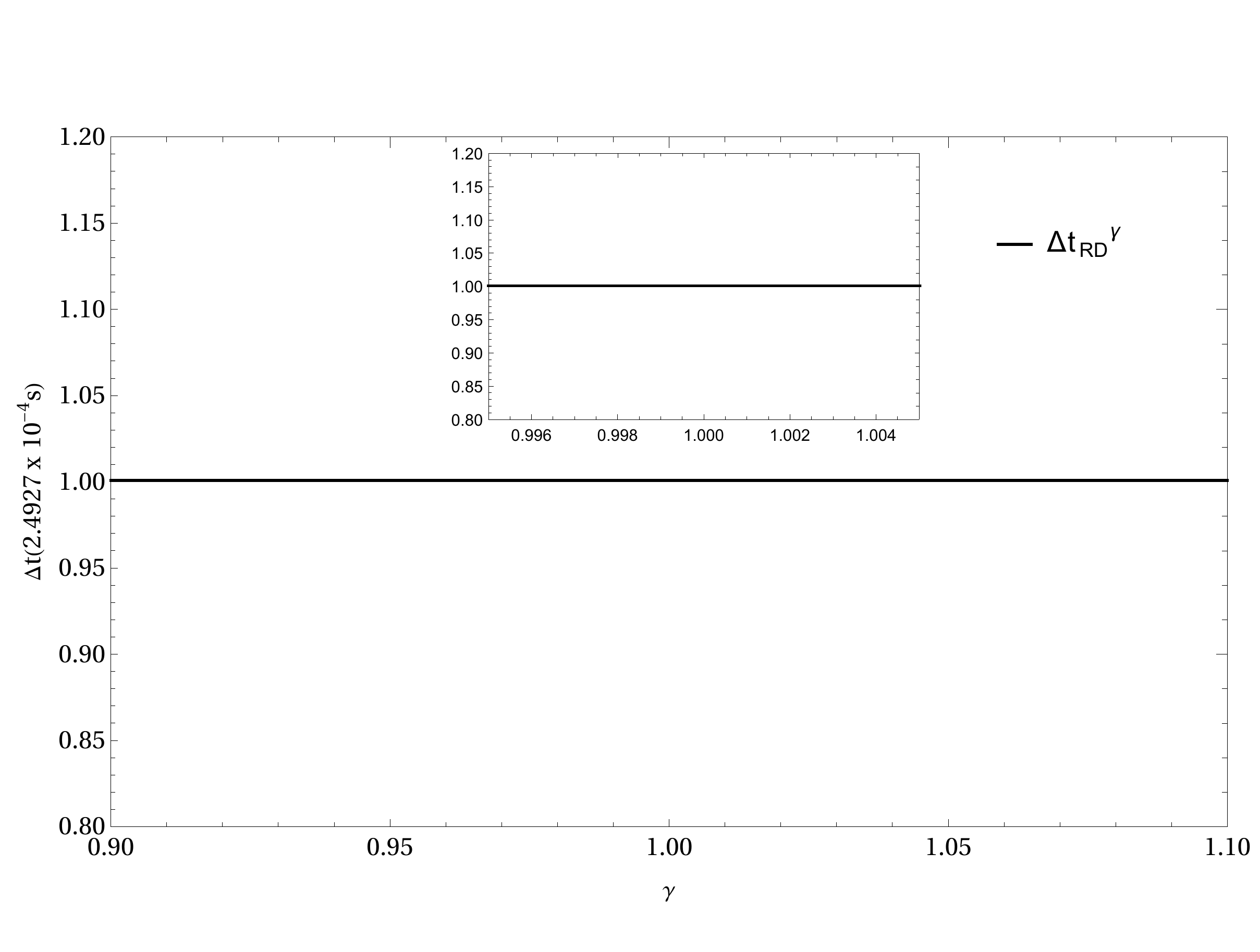}
    \end{center}
    \vspace{-0.5cm}
    \caption{ Numerical integration of Eq.~\eqref{eq-td-integral}. This Plot shows the time delay with respect to gamma when a radar signal travels back and forth through path close to the sun where the exterior of the sun assumed to be described by $ \gamma $ metric. \label{fig-timedelay}}
\end{figure*}

Let us now verify this in the strong-field regime. We numerically integrate Eq.~\eqref{eq-td-integral} without any approximations. The result of the integration is shown in Fig.~\ref{fig-timedelay}, which clearly shows that the Shapiro time delay in the $ \gamma $ metric is independent of the parameter $ \gamma $.

\section{Perihelion precession}\label{sec-pp}

To find the rate precession of perihelion in $ \gamma $-metric, let us now concentrate on the motion of massive particles. The metric is independent of $ t $ and $ \theta $ coordinates and hence we shall again have two timelike Killing vectors
\begin{equation}\label{eq-tdot-phidot}
    \begin{aligned}
        e &\equiv -\boldsymbol\varepsilon . \mathbf{v} = f^\gamma\left(\frac{dt}{d\tau}\right) \Rightarrow \left(\frac{dt}{d\tau}\right) = \frac{e}{f^\gamma}, \\
        l &\equiv \boldsymbol\eta . \mathbf{v} = r^2 f^{1-\gamma} \left( \frac{d\phi}{d\tau} \right) \Rightarrow \left( \frac{d\phi}{d\tau} \right) = \frac{l}{r^2f^{1-\gamma}}.  
    \end{aligned}
\end{equation}
A third constant of motion is given by the considering the requirement that the tangent vector to be timelike for massive particles
\begin{equation}
    \mathbf{v}.\mathbf{v} = g_{\alpha\beta}u^{\alpha}u^{\beta} = -1.
\end{equation}
On the equatorial plane, this equation can be written as
\begin{equation}
    -f^\gamma \dot{t}^2 + \left( \frac{f^{\gamma^2-\gamma}g^{1-\gamma^2}}{f} \right)\dot{r}^2 + f^{1-\gamma}r^2\dot{\phi}^2 = -1,
\end{equation}
where $ \dot{t} = dt/d\tau $, $ \dot{\phi} = d\phi/d\tau $ and $ \tau $ is the proper time. Now we can replace $ \dot{t} $ and $ \dot{\phi} $ using Eq.~\eqref{eq-tdot-phidot} and the equation becomes
\begin{equation}
    -f^{-\gamma}e^2 + \left( \frac{f^{\gamma^2-\gamma}g^{1-\gamma^2}}{f} \right)\dot{r}^2 + \frac{l^2}{r^2f^{1-\gamma}} = -1. 
\end{equation}
With some algebraic manipulations, this equation can be written as
\begin{equation}
    \dot{r}^2 + A(r)^{-1} \frac{l^2}{r^2f^{1-\gamma}} = A(r)^{-1} \left( f^{-\gamma}e^2 -1 \right),
\end{equation}
where
\begin{equation}
    A(r) = \frac{f^{\gamma^2-\gamma}g^{1-\gamma^2}}{f}.
\end{equation}
Now we employ a change of variable of the form $ r = 1/w $ and replace $ d/d\tau $ with $ (l w^2/f^{1-\gamma})d/d\phi $ and obtain
\begin{equation}
    \left( \frac{dw}{d\phi} \right)^2 + \frac{f^{1-\gamma}}{A(w)}w^2 = \frac{f^{2(1-\gamma)}}{l^2A(w)}\left( f^{-\gamma}e^2 - 1 \right). 
\end{equation}
Arranging the terms
\begin{equation}
    \left(\frac{dw}{d\phi}\right)^2 + w^2 = \frac{e^2}{l^2}X(w) - \frac{1}{l^2}Y(w) - w^2 Z(w), 
\end{equation}
where 
\begin{equation}
    \begin{aligned}
        X(w) &= \frac{f^{2-3\gamma}}{A(w)}, \\
        Y(w) &= \frac{f^{2(1-\gamma)}}{A(w)}, \\
        Z(w) &= \frac{f^{1-\gamma}}{A(w)} - 1.
    \end{aligned}
\end{equation}
By taking the derivative of the previous equation with respect to $ \phi $, we find
\begin{equation}
    \frac{d^2w}{d\phi^2} + w = F(w),
\end{equation}
which is the orbit equation. Here,
\begin{equation}
    F(w) = \frac{1}{2}\frac{dG(w)}{dw},
\end{equation}
and
\begin{equation}
    G(w) = w^2 Z(w) + \frac{e^2}{l^2}X(w) - \frac{1}{l^2}Y(w).
\end{equation}
Using the method as shown in \cite{Harko:2009qr,Capistrano:2019qdv}, we work with the previous orbit equation. The deviation angle $ \delta\phi $ can be found analytically using
\begin{equation}
    \delta\phi = \pi\frac{dF(w)}{dw}\Bigg|_{w_0},
\end{equation}
along with the constraint $ F(w_0) = w_0 $ for a near-circular orbit. 
In the weak field regime ($ Mw << 1 $), we can expand $ X(w) $, $ Y(w) $ and $ Z(w) $ in powers of $ w $ to write $ G(w) $ and $ F(w) $
\begin{equation}
    \begin{aligned}
        G(w) &= \frac{w^2 \left(3 \left(3 \gamma ^2-4 \gamma +1\right) e^2 M^2-3 (\gamma -1)^2 M^2\right)}{l^2} \\
        &+\frac{w \left(4 (\gamma -1) e^2 M-2 \gamma  M+4 M\right)}{l^2} \\
        &+\frac{e^2-1}{l^2}-(\gamma -1) (\gamma +1)
        M^2 w^4+2 M w^3 \\
        F(w) &= \frac{M \left(-\gamma +2 (\gamma -1) e^2+2\right)}{l^2} \\
        &+\frac{M w \left(3 (\gamma -1) (3 \gamma -1) e^2 M-3 (\gamma -1)^2 M\right)}{l^2}\\
        &-2 (\gamma -1) (\gamma +1) M^2 w^3+3 M w^2
   \end{aligned}
\end{equation}
So the circular orbits will be given by the roots of the equation $ F(w_0) = w_0 $. Explicitly 
\begin{equation}
    A_0 + A_1 w_0 + A_2 w_0^2 + A_3 w_0^3 = w_0,
\end{equation}
where 
\begin{equation}
    \begin{aligned}
        A_0 &= \frac{M \left(-\gamma +2 (\gamma -1) e^2+2\right)}{l^2}, \\
        A_1 &= \frac{M^2 \left[3 (\gamma -1) (3 \gamma -1) e^2 -3 (\gamma -1)^2\right]}{l^2}, \\
        A_2 &= 3M, \\
        A_3 &= -2 (\gamma -1) (\gamma +1) M^2.
    \end{aligned}
\end{equation}
Neglecting higher order corrections, the solution to the equation $ F(w_0) = w_0 $ can be written as
\begin{equation}
    w_0 = \frac{A_0}{1-A_1},
\end{equation}
which reduces to $ w_0 = M/l^2 $ for $ \gamma = 1 $. Using the above results, the perihelion precession $ \delta \phi $ can be written as
\begin{equation}\label{eq-perih-res}
    \begin{aligned}
        \delta \phi &= \pi \frac{dF(w)}{dw}\Bigg|_{w_0} \\
        &= \pi \Bigg[\frac{M^2 \left[3 (\gamma -1) (3 \gamma -1) e^2 -3 (\gamma -1)^2 \right]}{l^2} \\
        & \ \ \ \ \ \ \ \ \ \ -6 (\gamma -1) (\gamma +1) M^2 w_0^2+6 M w_0\Bigg].
    \end{aligned}
\end{equation}

For $ \gamma = 1 $, we recover the classical General Relativistic result $ \delta\phi_{\rm GR} = 6\pi\left(M^2/l^2\right) $.

\begin{figure*}[]
    \begin{center}
        \includegraphics[width=8.5cm]{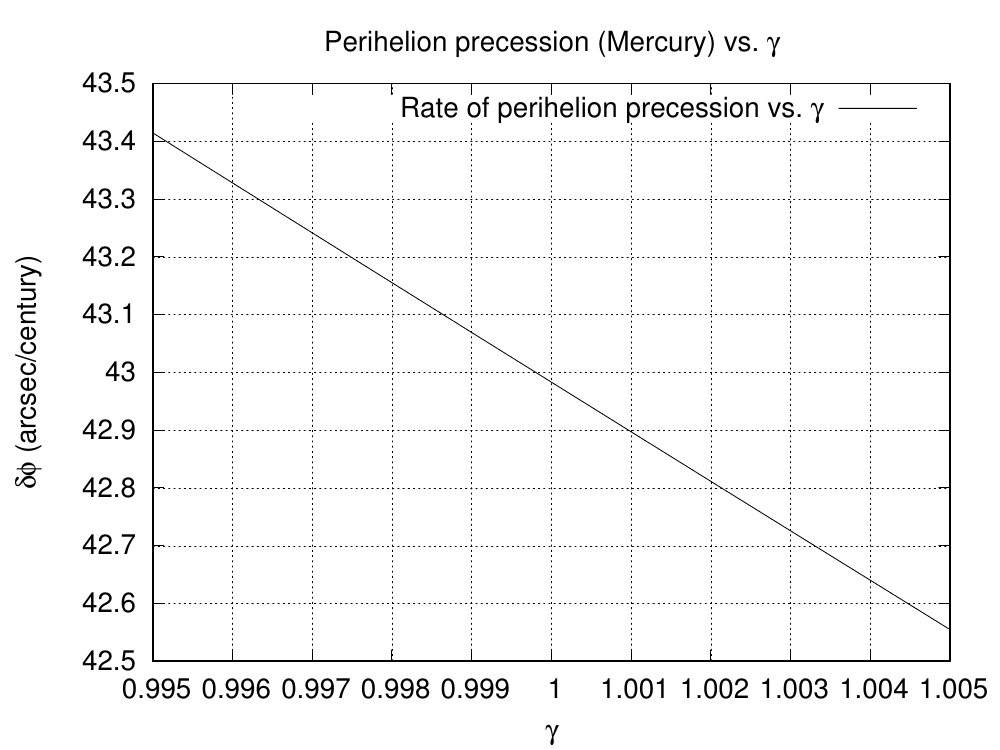}
        \includegraphics[width=8.5cm]{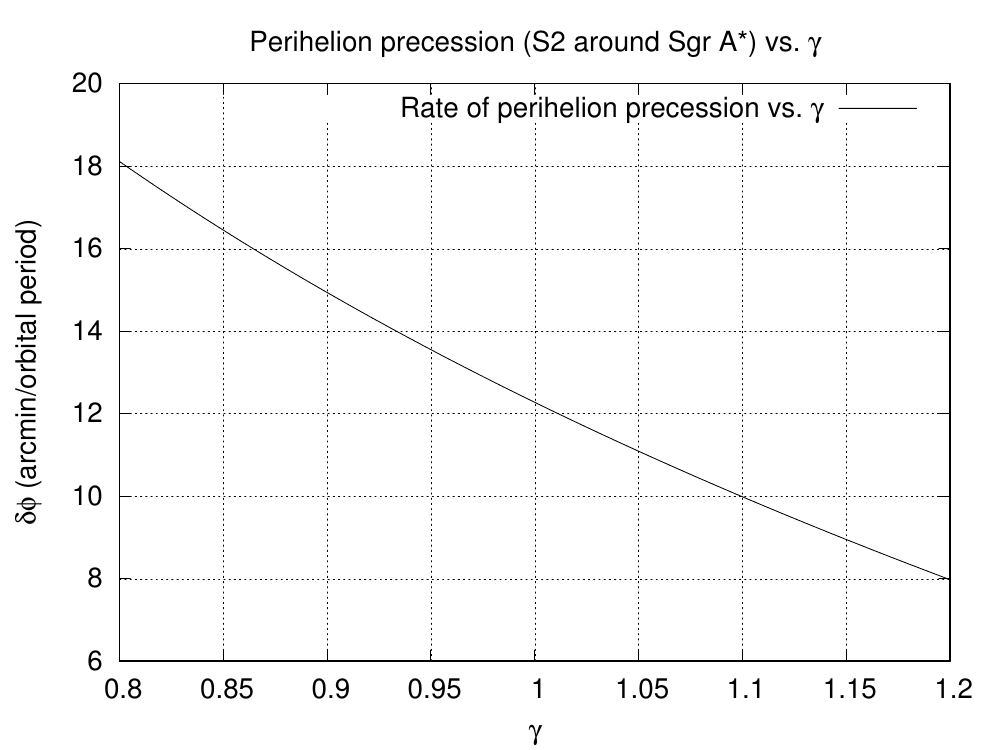}
    \end{center}
    \vspace{-0.5cm}
    \caption{ Perihelion shift as a function of the deformation parameter for Mercury (left) and S2 star (right) with respect to $ \gamma $.  \label{perih}}
\end{figure*}

For the $ \gamma $ metric, we report the variation of perihelion precession angle of Mercury around the sun (left panel) and S2 star around Sgr A* (right panel) with respect to the deformation parameter $ \gamma $ in Figure \ref{perih}. In Table~\ref{tab2}, we show the numerical estimates of perihelion precession for Mercury for values of $ \gamma = 0.95, 0.97, 1.0, 1.02 $ and $ 1.05 $. We can put constraints on the parameter $ \gamma $ in the solar system using the measured values of perihelion precession for Mercury $ \delta \phi_{\rm prec} $. The observed value of the perihelion shift of Mercury is $ 42.98 \pm 0.04 $ arcsec/century \cite{Will:2005yc,Will:2014kxa}. Using this data, $ \gamma $ can be constrained at $ 1.0000 \pm 0.0005 $ where we clearly recover the Schwarzschild limit. On the other hand, for S2, the measured value of the orbital precession per orbit is \cite{GRAVITY:2020gka} $ \delta \phi_{\rm prec} = 12.1 \times (1.10 \pm 0.19 ) $ arcminute/orbit. This constrains $ \gamma $ at $ 0.96 \pm 0.08 $ for Sgr A* on the plane of the orbit in the weak field limit. We still recover the Schwarzschild limit. However, note that this analysis is preliminary. For better constraints, full numerical simulations with all orbital parameters should be done. With the increase in precision of the measured value in the coming years, we would be able to obtain a more stringent constraint on the deformation parameter. Similarly, an analysis in the strong field regime would also require full numerical simulations of the orbit equations, which then can be applied to the motion of S stars to constraint deviation from spherical symmetry.

\begin{table}[t]
\centering
\begin{tabular}{ |p{2cm}|p{3cm}| } 
 \hline
 $\gamma$ & $\delta \phi$ (arcsec/century) \\
 \hline
 0.95 & 47.45 \\ 
 0.97 & 45.62 \\
 1.0 & 42.98 \\
 1.02 & 41.28 \\
 1.05 & 38.83 \\
 \hline
\end{tabular}
\caption{Estimates of precession of perihelion of Mercury for different values of $\gamma$.}\label{tab2}
\end{table}

\section{Outlook and discussion}\label{sec-od}

In the present work, we tested the applicability of the $ \gamma $-metric for astrophysical observations. We studied strong and weak gravitational lensing, radar echo delay and perihelion precession in the context of the considered metric spacetime. We demonstrated that most of the weak field experiment observables do not show any dependence on $ \gamma $, except perihelion precession. However, strong lensing shows a significant effect on light bending in $ \gamma $-metric.  

We started with gravitational lensing in the weak field limit and calculated the deflection angle of photons. The expression of the deflection angle depends on the deformation parameter $ \gamma $ and the parameter related to the mass $ M $. This dependence comes in the form $ \gamma M $ which is the observable mass of the system. Therefore, it is unlikely to independently constrain $ \gamma $ with weak lensing experiments. On the other hand, the mechanism in strong lensing is quite involved and we can see a clear dependence of the observational quantities in the deformations parameter. This can be seen in Fig.~\ref{fig-sl}. The deflection angle in strong lensing increases with decreasing $ \gamma $ and blows up at $ \gamma = 0.5 $. This is because of the disappearance of the photon sphere at the corresponding value of $ \gamma $.  

We then concentrated on two classic solar system experiments, namely Shapiro delay and perihelion precession. We showed that light rays in the time delay experiment are not affected by the deformation parameter. Finally, we studied the precession of perihelion in this spacetime and showed that perihelion shift depends on the deformation parameter, i.e. with increasing $ \gamma $ the shift in perihelion decreased as can be seen in Fig.~\ref{perih}. Using the currently accepted value of the perihelion shift of Mercury, we obtained numerical constrain on the deformation parameter which is $ \gamma = 1.0000 \pm 0.0005 $. Similarly, data from S2 star contraints $ \gamma $ at $ 0.96\pm0.08 $. Shapiro time delay is still a weak field effect in the solar system and it is expected that the deformation parameter of a modified spacetime will affect the travel time of a signal if the signal passes close to the massive object. However, in our analysis of $ \gamma $-metric, we can see that the integrand of the time delay integral does not depend on the non-trivial $ g_{\phi\phi} $ term of the metric and the contribution from $\gamma$ in $ g_{tt} $ and $ g_{rr} $ seems to contribute only towards the observable mass of the massive object. On the other hand, in strong lensing and perihelion precession, the deviation from the Schwarzschild value comes from the $g_{\phi\phi}$ term.

Finally, concerning the possibility of constraining deviations from spherical symmetry through observation of gravitational lensing, our results show that only precise measurements of observables in strong field lensing would allow us to distinguish a black hole from the $ \gamma $-spacetime metric. On the other hand, more precise observation of the perihelion shift of Mercury and S stars in the future would allow us to put much tighter bounds on $ \gamma $ in the respective systems.

\noindent
\acknowledgements
HC thank Prof. Bobomurat Ahmedov, Dr. Daniele Malafarina and Dr. Ahmadjon Abdujabbarov for useful discussions. HC also acknowledges support from University of Chinese Academy of Sciences through Special Research Assistant fellowship (SRA). YT is supported by Natural Science Foundation of China (NSFC) under Grants No.~11851302, the Fundamental Research Funds for the Central Universities and Key Research Program of the Chinese Academy of Sciences, Grant No. XDPB15.

\appendix

\section{Deflection angle in the strong field limit: Bozza's method}\label{appA}

To find the coefficients of the deflection angle in Eq.~\eqref{sl-def-ang}, first two new variables are defined as
\begin{equation}
    \begin{aligned}
    y &= A(r) \\
    z &= \frac{y - y_0}{1 - y_0}
    \end{aligned}
\end{equation}
where $ y_0 = A_0 = A(r_0) $. In terms of these new variables the integral in deflection angle \eqref{sl-int} becomes 
\begin{equation}\label{Randf}
    \begin{aligned}
    &I(r_0) = \int_0^1 R(z,r_0)f(z,r_0)dz, \\
    &R(z,r_0) = \frac{2\sqrt{By}}{CA'}(1-y_0)\sqrt{C_0}, \\
    &f(z,r_0) = \frac{1}{\sqrt{y_0-[(1-y_0)z+y_0]\frac{C_0}{C}}}
    \end{aligned}
\end{equation}
where all functions without the subscript $ 0 $ are evaluated at $ x = A^{-1}[(1-y_0)z+y_0] $.

Here the function $ R(z,r_0) $ is regular for all values of $ z $ and $ r_0 $ but $ f(z,r_0) $ diverges for $ z \rightarrow 0 $. The argument under the square root in $ f(z,r_0) $ is expanded to the second order in $ z $,
\begin{equation}
    \begin{aligned}
    f(z,r_0) \sim f_0(z,r_0) = \frac{1}{\sqrt{\alpha z + \beta z^2}},
    \end{aligned}
\end{equation}
where,
\begin{equation}\label{alphabeta}
    \begin{aligned}
    \alpha &= \frac{1-y_0}{C_0A_0'}(C_0'y_0 - C_0A_0'), \\ 
    \beta &= \frac{(1-y_0)^2}{2C_0^2A_0'^3}\Big[ 2C_0C_0'A_0'^2 + \left( C_0C_0'' - 2C_0'^2 \right)y_0A_0' \\
    &- C_0C_0'y_0A_0'' \Big].
    \end{aligned}
\end{equation}
In case of a non-zero $ \alpha $, the leading order divergence is $ z^{1/2} $, which can be integrated to give a finite result. The integral diverges when $ \alpha = 0 $, since the leading order divergence is $ z^{-1} $. Analyzing the form of $ \alpha $, we can see that it vanishes for $ r_0 = r_m $, resulting in capture of each photon having $ r_0 < r_m $. 

Now to solve the integral, it is split into two parts, $ I_D $ (contains the divergence) and $ I_{R} $ (regular):
\begin{equation}
    \begin{aligned}
    I(r_0) &= I_D(r_0) + I_R(r_0) \\
    I_D(r_0) &= \int_0^1 R(0,r_m)f_0(z,r_0)dz, \\
    I_R(r_0) &= \int_0^1 g(z,r_0)dz,
    \end{aligned}
\end{equation}
where
\begin{equation}
    g(z,r_0) = R(z,r_0)f(z,r_0) - R(0,r_m)f_0(z,r_0).
\end{equation}
We can see that $I_R(r_0)$ is the original integral with the divergence subtracted. Now both the integrals should be solved separately and then summed up rebuilding the deflection angle. 

Let us first handle the integral $ I_D(r_0) $. It can be solved exactly. The results is
\begin{equation}
    I_D(r_0) = R(0,r_m)\frac{2}{\sqrt{\beta}}\log \frac{\sqrt{\beta}+\sqrt{\alpha+\beta}}{\sqrt{\alpha}}.
\end{equation}
We are interested only in the terms up to $ \mathcal{O}(r_0-r_m) $, so we expand $ \alpha $ as 
\begin{equation}
    \alpha = \frac{2\beta_m A_m'}{1-y_m}(r_0-r_m) + \mathcal{O}(r_0 - r_m)^2,
\end{equation}
where 
\begin{equation}\label{betam}
    \beta_m = \frac{C_m(1-y_m)^2(C_m''y_m-C_mA_m'')}{2y_m^2C_m''}.
\end{equation}
We then substitute it in $ I_D(r_0) $
\begin{equation}
    I_D (r_0) = -a \log\left( \frac{r_0}{r_m} - 1 \right) + b_D + \mathcal{O}(r_0 - r_m)
\end{equation}
where
\begin{equation}
    \begin{aligned}
    &a = \frac{R(0,r_m)}{\sqrt{\beta_m}}, \\
    &b = \frac{R(0,r_m)}{\sqrt{\beta_m}}\log \frac{2(1-y_m)}{A_m'r_m}.
    \end{aligned}
\end{equation}
$ I_D(r_0) $ yields logarithmic leading order divergence of the deflection angle. 

Now let us concentrate on the regular term. To find $ b $ we need to add an analogous term coming from the regular part of the integral to $ b_D $. First we expand $ I_R(r_0-r_m) $
\begin{equation}
    I_R(r_0) = \sum_{n=0}^{\infty} \frac{1}{n!}(r_0-r_m)^n\int_0^1 \frac{\partial^n g}{\partial r_o^n}\Bigg|_{r_0 = r_m}dz
\end{equation}
and evaluate the single coefficients. 

Not subtracing the divergent part would have resulted in an infinite coefficient for $ n = 0 $, but the other coefficients would be finite. However, $ g(z,r_0) $ is regular for $ z = 0 $ as $ r_0 \rightarrow r_m $. We are only interested to terms up to $ \mathcal{O}(r_0 - r_m) $, so just retaining the leading order term
\begin{equation}
    I_R(r_0) = \int_0^1 g(z,r_m)dz + \mathcal{O}(r_0-r_m).
\end{equation}
Then we have 
\begin{equation}\label{bR}
    b_R = I_R(r_m)
\end{equation}
which is the term needed to be added to $ b_D $ to get the regular coefficient. So finally,
\begin{equation}
    b = -\pi + b_D + b_R. 
\end{equation}
The term $ b_R $ can be evaluated numerically for all metric forms since the integrand does not contain any divergences. However, for Schwarzschild metric, it is solved exactly. For $ \gamma $-metric we expand the integrand over $ (\gamma - 1) $ and try to obtain an analytical expression. 

Now to go from $ \alpha (r_0) $ to $ \alpha (\theta) $ we expand the equation for impact parameter
\begin{equation}\label{um}
    \begin{aligned}
    &u_m = \sqrt{\frac{C_m}{y_m}}, \\
    &u-u_m = c(r_0 - r_m)^2,
    \end{aligned}
\end{equation}
where 
\begin{equation}
    c = \frac{C_m''y_m - C_mA_m''}{3\sqrt{y_m^3C_m}} = \beta_m\sqrt{\frac{y_m}{C_m^3}}\frac{C_m'^2}{2(1-y_m)^2}.
\end{equation}
Using this relation, the deflection angle can be expressed as a function of $ \theta $
\begin{equation}
    \alpha(\theta) = -\Bar{a}\log\left( \frac{\theta D_{OL}}{u_m} - 1 \right) + \Bar{b}
\end{equation}
where
\begin{equation}\label{abarbbar}
    \begin{aligned}
    &\Bar{a} = \frac{a}{2} = \frac{R(0,r_m)}{2\sqrt{\beta_m}} \\
    &\Bar{b} = -\pi + b_R + \Bar{a} \log \frac{2\beta_m}{y_m}.
    \end{aligned}
\end{equation}
So finally, in order to calculate the deflection angle as a function of $ \theta $, we need to 
\begin{itemize}
    \item solve Eq.~\eqref{psphere} to find $ r_m $.
    \item write $ \beta_m $ and $ R(0,r_m) $ from Eq.~\eqref{betam} and Eq.~\eqref{Randf} respectively
    \item compute $ b_R $ from Eq.~\eqref{bR} by a an expansion of the parameters of the metric
    \item compute $ u_m $, $ \Bar{a} $ and $ \Bar{b} $ from Eq.~\eqref{um} and \eqref{abarbbar} respectively. 
\end{itemize}
The only integral involved in the process is calculating $ b_R $.

\end{document}